\documentclass[pra,superscriptaddress,twocolumn,english, longbibliography, showpacs=false, nofootinbib]{revtex4-2}

\usepackage[T1]{fontenc}
\usepackage[utf8]{inputenc}
\usepackage{xcolor}
\usepackage{babel}
\usepackage{amsmath}
\usepackage{amssymb}
\usepackage{subfigure}
\usepackage{graphicx}
\usepackage{physics} 
\usepackage{dsfont}
\usepackage{hyperref}
\usepackage[colorinlistoftodos, color=green!40, prependcaption]{todonotes}
\usepackage{amsthm}
\usepackage{mathtools}
\usepackage[left=23mm,right=13mm,top=35mm,columnsep=15pt]{geometry} 
\usepackage{adjustbox}
\usepackage{placeins}
\usepackage{csquotes}
\usepackage{mathrsfs}
\usepackage{tikz}
\usetikzlibrary{quotes,angles}

\begin{document}

\title{Analogue gravitational lensing in Bose-Einstein condensates}

\author{Decheng Ma}
\affiliation{Department of Physics, Anshan Normal University, Anshan 114056, China}

\author{Chenglong Jia}
\affiliation{Lanzhou Center for Theoretical Physics, Key Laboratory of Theoretical Physics of Gansu Province, and Key Laboratory of Quantum Theory and Applications of MoE, Lanzhou University, Lanzhou, Gansu 730000, China}
\affiliation{Key Laboratory for Magnetism and Magnetic Materials of the MoE, Lanzhou University, Lanzhou 730000, China}

\author{Enrique Solano}
\affiliation{Kipu Quantum, Greifswalderstrasse 226, 10405 Berlin, Germany}

\author{Lucas C. Céleri}
\email{lucas@qpequi.com}
\affiliation{QPequi Group, Institute of Physics, Federal University of Goi\'as, Goi\^ania, Goi\'as, 74.690-900, Brazil}

\begin{abstract}
We consider the propagation of phonons in the presence of a particle sink with radial flow in a Bose--Einstein condensate. Because the particle sink can be used to simulate a static acoustic black hole, the phonon would experience a considerable spacetime curvature at appreciable distance from the sink. The trajectory of the phonons is bended after passing by the particle sink, which can be used as a simulation of the gravitational lensing effect in a Bose--Einstein condensate. Possible experimental implementations are discussed.
\end{abstract}

\maketitle

\section{Introduction}
\label{Intro}

Analogue gravity has been considered and developed since 1923~\cite{Gordon1923,Felice1971}. However, it was the discovery of a formal analogy between the dynamics of massless scalar fields in black hole spacetimes and sound waves in a moving fluid~\cite{Unruh1981} that brought about the dawn of modern gravitational analogues. The~importance of this result relies on the recognition that Hawking radiation is not related to general relativity itself, instead~being is an effect that must be present whenever we consider quantum field theories in curved spacetimes with an event horizon. Subsequently, the~field has witnessed impressive development both theoretically and experimentally, expanding beyond sonic waves in classical fluids. Among~these achievements, we can mention phonons in Bose--Einstein condensates~\cite{Lahav2010,Garay2000,Yatsuta2020,Nova2019,Steinhauer2014,Steinhauer2016,Kolobov2021,Steinhauer2017,Fedichev2003,Fedichev2004,Eckel2018,Vieira2023,Balbinot2022}, surface waves on water flows~\cite{Rousseaux2008,Weinfurtner2011,Peloquin2016,Torres2017,Chakraborty2020}, slow light in optical systems~\cite{sheng2013,Tinguely2020,Bekenstein2017,Drori2019,Sheng2018,Zhong2018,Brevik2019,Brevik2020}, and magnons in magnetic systems~\cite{Jannes2011,Molina2017}. These analogue systems have become an important testing ground for certain aspects of curved spacetimes, especially those involving quantum systems where the superradiance effect~\cite{Torres2017,Basak2003}, Hawking radiation~\cite{Yatsuta2020,Nova2019,Steinhauer2014,Steinhauer2017}, and cosmological particle production~\cite{Fedichev2003,Fedichev2004,Eckel2018} are present. Furthermore, analogue gravity can provide insights into the quantum nature of gravity while aiding in understanding the spacetime back-reaction~\cite{Balbinot2005,Patrick2021} and the cosmological constant problem~\cite{Finazzi2012}. Moreover, such models can be understood in terms of emergent gravity, which may help us in the search for a quantum theory of gravity~\cite{Volovik2003,Barcelo2001}.

Among the phenomena that can be mimicked is gravitational lensing, which is a general name for light deflection by a gravitational field generated by astrophysical sources such as stars, black holes, and galaxies~\cite{Carroll2019}. The~first observation of this effect was carried out in 1919, when Eddington's team confirmed Einstein's prediction of the deflection of light by the gravitational field produced by the sun~\cite{Dyson1920}. Since then, many other theoretical and experimental developments have appeared~\cite{Bartelmann2010,Chakraborty2022}.

Considering acoustic propagation around a vortex in a superfluid system, it has been shown that the phonon trajectories are deflected by an angle determined by the vortex circulation~\cite{Fischer2002,Fischer2003}. In~this way, the~vortex acts as a converging lens for the phonons trajectories. Light deflection was observed in~\cite{sheng2013}, where a microstructured waveguide around a microsphere was employed in order to mimic a curved spacetime. Optical materials were employed to simulate equatorial Kerr--Newman null geodesics, allowing the study of light trajectories in a non-static spacetime~\cite{Tinguely2020}.

Here, we consider optical Bose--Einstein condensates (BEC) to build a simulation of the gravitational lensing phenomenon. When particles of integer spin accumulate in the ground state at high density and low temperature, a~BEC is formed~\cite{Leggett2001}. However, in~the case of massless particles this is more complicated, as lowering the temperature of a gas consisting of massless particles does not preserve the number of particles. In~fact, the~ground state, or~vacuum, has no particles at all. This problem can be circumvented by considering a dye-filled optical cavity~\cite{Klaers2010a}, where a photonic BEC can be achieved~\cite{Klaers2010b}. The~process works because the system effectively provides a mass to the photons, which can then thermalize to the cavity ground state while conserving the number of particles. The~physical process behind this is the multiple scattering of the trapped photons by dye molecules, ultimately leading to~thermalization.

Specifically, we consider phonons propagating in a photonic BEC using the particle sink model proposed in~\cite{Liao2019}, which has been used to simulate Schwarzschild black holes and Hawking radiation. Because such structures create an effective curved spacetime, we expect to observe its influence on the trajectories of the phonons. We show that the quasiclassical scattering process of phonons by a particle sink leads to a nonvanishing scattering angle, deviating the trajectories from a straight line. In~other words, we show that the effect of the effective spacetime curvature on the phonon trajectory, as~predicted by the gravitational lensing effect, is the same as a converging lens. This analogue gravitational lensing effect may provide direct evidence for the curved nature of the acoustic spacetime. In addition, we discuss a possible experimental implementation of this effect in a photonic BEC by considering the angle-resolved photoluminescence spectrum method as a suitable and available technique. This can be used to track the photon trajectory and measure the expected deviation angle, which can be compared with the predictions based on the gravitational lensing~effect.

The advantage of this approach is that photonic condensates are formed at room temperature, contrary to the very low temperatures that are employed in other systems. Moreover, due to the simplicity of the experimental setup in which such condensates are implemented, namely, a~dye-filled pumped optical cavity, this approach offers a manifold of possibilities for analogue~gravity.

We proceed as follows. A~description of the system under consideration and how it can be employed to simulate a curved spacetime is presented in Section~\ref{sec:analogue}. Section~\ref{sec_lens} is devoted to presenting our main result, the~analogue gravitational lensing in a photonic BEC. We close the paper in Section~\ref{sec_conclusion} which, in addition to~our final conclusions, contains several open questions and possible routes for future~investigation.

\section{Analogue~Metric}\label{sec_metric}
\label{sec:analogue}

In a dye-filled cavity, photons can reach the equilibrium state at room temperature due to scattering with the dye molecules. Beyond~providing a confining potential to trap the photons, the~cavity additionally leads to a non-vanishing effective mass for the photons, making them behave as a two-dimensional trapped massive bosonic system~\cite{Klaers2010a}. This effect makes the existence of the photonic condensate possible. The~photon chemical potential can be freely adjusted using the fluorescence-induced thermalization, while the photon density can be tuned using the pumping light intensity. Upon~increasing the photon density, when the photon number exceeds the critical number, the~photonic BEC transition can be induced in the cavity, with macroscopic photons settling into the lowest energy state~\cite{Klaers2010a,Klaers2010b}.

Considering Maxwell's equations in a nonlinear dielectric medium (dye-filled cavity), the~dynamical equation for the wave function describing the condensate photon gas (including incoherent pumping) is, in~the mean-field approximation, the~effective dissipative Gross--Pitaevskii equation (GPE)~\cite{Nyman2014}:
\begin{equation}
i\hbar \pdv{\psi}{t} = \left[V(\mathbf{r})-\frac{\hbar^2}{2m}\nabla_{\bot}^2+g|\psi|^2+i(\gamma_{net}-\Gamma|\psi|^2)\right]\psi.
\label{eq:GP}
\end{equation}
In 
 this equation, the~net dissipation (the difference between the pumping and dissipation rates of the cavity) is described by $\gamma_{net}$, the~constant $\Gamma$ assures stability in the sense that the last term in the brackets vanishes in~the steady state ($\Gamma = \gamma_{net}/|\psi (0)|^2$), $g$ denotes the effective inter-particle interaction strength arising from the Kerr nonlinearity of the system, and $V(\mathbf{r})$ is the effective potential energy which includes the detuning between the cavity mode and pumping field. The~number of particles in the condensate can be computed in terms of the photonic wave function $\psi$ as $N_{\rm BEC}=\int \dd\xi \, \psi(\xi)|^2$, where $\xi$ is the considered set of~coordinates.

The effective photon mass $m$ is determined by the relation $\hbar\omega=mc_L^2/n_L$, where $\omega$ is light angular frequency, $n_L$ is the refractive index, and $c_L$ denotes the speed of light. In~the steady state limit, the~decay rate and pumping rate compensate each other;~the condensate density is provided by $n_{0}=\gamma_{net}/\Gamma$. In particular, as~demonstrated by Kneer~\cite{Kneer1998}, the~condensate can reach a stationary state when the pumping term is beyond a typical threshold, which can be described by the conventional GPE without the pumping and dissipation terms. In~the same spirit, a stable dissipative vortex structure can be created in exciton--polariton condensates~\cite{Solnyshkov2019,Jacquet2020}. From~here on, we assume that the condensate of light is in the steady state; hence, the dissipative term in Equation~\eqref{eq:GP} can be safely neglected. Unlike the exciton polariton BECs, the~photonic BECs are in the weak light--matter coupling limit, and~are usually very close to the thermal equilibrium state~\cite{Klaers2010a}. The~decay processes in photonic BECs are negligibly small, and~the photon number is conserved. Being~low energy excitations in the condensate, the~phonons are stable quasiparticles with a very long lifetime~\cite{Nyman2014}, allowing them to be measured, as~we discuss~latter.

Moreover, the~dissipation and pumping effects can have important influences on the condensate. With~localized dissipative perturbations in atomic BEC, the~phase and amplitude of the condensate can be engineered and nonlinear coherent excitations can be  generated~\cite{Brazhnyi2009}. When an atomic BEC is coupled to a sources of uncondensed atoms and a sink, the~dissipative BEC obeys a complex Ginzburg--Landau equation and~the density of the condensate displays nontrivial spacetime correlations~\cite{Arecchi2000}. In~the photonic BEC with exciton--polariton quasiparticles pumped in an annular geometry, multiple charged vortex states can be spontaneously generated during the condensate formation and can be stabilized by pumping fluxes~\cite{Alperin2021}.

Creating a stable analogue black hole structure in a condensate is challenging. For~atomic condensates, any sink-like structure can quickly lead to depletion of the condensate. A~solution to this problem is to switch to different metastable quasi-particle systems with external pumping to compensate for the loss of particles such as polaritons or photons~\cite{Carusotto2013,Solnyshkov2019,Klaers2010b,Liao2019,Klaers2011,Walker2018,Nyman2014}. In~order to create a stable analogue Schwarzschild black hole in a photonic condensate, a~particle sink with a velocity singularity in the center (created by a sink in the cavity trapping the photons) was proposed in~\cite{Liao2019}. Because~of the sink in the system, the~photons leak out of the cavity, inducing a radial flow towards the sink. Therefore, we expect a particle sink with the velocity profile provided by~
\cite{Liao2019}
\begin{equation}
\textbf{v} = -\xi c\frac{c_{0}}{r}\hat{r},
\end{equation}
where $c=\sqrt{gn_0/m}$ is the sound velocity, $\textbf{r}$ is the radial coordinate, and~$\xi=\hbar/mc$ stands for the healing length of the condensate. In~addition, $c_{0}>0$ denotes a constant parameter, which is determined by the rate of particles extracted from the condensate by the sink. Below, we associate ${c_0}^2$ with the black hole mass. In~contrast to the usual vortex in an irrotational fluid or BEC, the~particle sink is not rotating, and as such its velocity does not have any azimuthal component. This static particle sink can be mapped into an effective metric which is an analogue of a two-dimensional Schwarzschild black hole~\cite{Liao2019}. We describe this construction in the following paragraphs. The~idea here is to create a quantum fluid where sound (phonons) can travel. The~above velocity profile leads to subsonic and supersonic regions in the fluid. The~transition between them is the analogue of the event horizon, in~the sense that sound cannot leave the supersonic~region.

We are interested here in the dynamics of small perturbations around a stationary background characterized by the particle density and the velocity profile. Within~the valid regime of hydrodynamics, the~perturbations for the density and the phase obey a set of two coupled differential equations, called a~Bogoliubov system. This system can be transformed into a single differential equation for the phase perturbation $\theta$~\cite{Gross1961,Pitaevskii1961,Garay2001,Liao2019}
\begin{equation}
    \left\lbrace\nabla^2-\left(\partial_t-\frac{c_0}{r}\hat{\mathbf{r}}\cdot\nabla\right)^2\right\rbrace\theta(\mathbf{r},t)=0 \, ,
\end{equation}
where for~simplicity we have defined the dimensionless variables $r\rightarrow\xi r$ and $t\rightarrow \xi t/c$.

Now, in~order to make contact with the physics of curved spacetimes, we introduce an effective acoustic metric $g_{\mu\nu}$. By~doing this, the~previous equation can be recast as a wave equation on a curved spacetime:
\begin{equation}
\label{Kleinwave}
  \frac{1}{\sqrt{-g}}\partial_{\mu}(\sqrt{-g}g^{\mu\nu}\partial_{\nu})\theta = 0 , 
\end{equation}
with the effective acoustic metric provided by
\begin{equation}
 \dd s^2=-f(r)\dd\tau^2+f^{-1}(r)\dd r^2+r^2\dd\varphi^2.
 \label{eq_metric}
\end{equation}
In this equation, $f(r) = 1-c_0^{2}/r^{2}$ is the warp factor. To~obtain this form of the metric, we have applied the following  coordinate transformation:
\begin{equation}
\dd \tau = \dd t - \frac{c_0}{r(1-\frac{c_0^{2}}{r^{2}})}\dd r.
\end{equation}

The metric in Equation~\eqref{eq_metric} is written in similar form to the Schwarzschild metric when restricted to the equatorial plane of the gravitational black hole, with~the parameter $c_0^2$ playing the role of the black hole mass. Because~the photonic BEC is two-dimensional, we can only simulate certain properties of Schwarzschild black holes in the equatorial plane, and~the effective acoustic metric is only similar to the Schwarzschild metric, not identical. Furthermore, in~the derivation of the effective acoustic metric, we have dropped a conformal factor. In~the analogue gravity scenario, a~Schwarzschild black hole usually cannot be perfectly simulated with the acoustic metric, which is only related to the Schwarzschild black hole through an extra conformal factor~\cite{Vieira2023,Bilic2022,Oliveira2021,Visser1998}. However, because the phonon trajectories are conformal-invariant, these factors have no influence on the phonon trajectories and can be safely neglected~\cite{Barcelo2011,Visser1998}. Although~we cannot perfectly reproduce the Schwarzschild metric in the photonic BEC, the~effective metric is rotationally symmetric and has a warp factor describing the spacetime curvature, sharing many similarities with the Schwarzschild metric. An~important difference between the acoustic metric and the Schwarzschild one is that the latter is not a solution of Einstein's field equation; instead, it is determined by the hydrodynamical equations of the system. Furthermore, we must recall that in the absence of a cosmological constant, in~a $2+1$ dimensional spacetime, the~curvature tensor for~the vacuum Einstein field equations is zero and the spacetime is locally flat. Hence, there exists no black hole solution of the vacuum in Einstein's equation in a $2+1$ dimensional spacetime with zero cosmological constant~\cite{Padmanabhanbook}. However, with~this metric we can simulated many kinetic properties of Schwarzschild black holes in the equatorial plane, such as the Hawking radiation around the horizon and, as~demonstrated here, the~gravitational lensing~effect.

The event horizon associated with the metric in Equation~(\ref{eq_metric}) appears when the radial flow velocity towards the center exceeds the speed of sound, specified by the metric singularity at
\begin{equation}
    r_h = c_0.
\end{equation}

Inside the event horizon ($r<r_h$), the~light fluid moves inwards at supersonic speed, and~any phase fluctuations behind the event horizon are swept into the acoustic black hole without escaping the event horizon. Therefore, the~acoustic event horizon is a surface that splits the space into two regions, a~subsonic (external) and a supersonic (internal) one, playing the same role in the analogue model as the gravitational horizon in a black~hole.

Note that we have a black hole solution in $2+1$ dimensions, which is not present in general relativity. This is a consequence of the fact that Einstein's equations do not play a role in the analogue system described here. In~order to obtain an idea of the curvature of our effective spacetime, we can compute the Kretschmann scalar, which is provided by
\begin{equation}
    K = R_{\alpha\beta\gamma\delta}R^{\alpha\beta\gamma\delta} = \frac{44c_{0}^{2}}{r^8},
\end{equation}
with $R_{\alpha\beta\gamma\delta}$ being the Riemann curvature tensor. Here, $K$ is well behaved at $r_h$, showing that the singularity at $r_h$ is just a breakdown of the employed coordinate system and not a true singularity. The~genuine singularity of this acoustic spacetime lies at $r=0$, which is at the center of the particle~sink.

In addition, the~Ricci scalar of the metric is straightforward to compute:
\begin{equation}
R_c = g^{\mu\nu}R^{\alpha}_{\mu\alpha\nu} = \frac{2c_{0}^{2}}{r^4}.
\end{equation}
The Ricci scalar characterizes the curvature of the acoustic spacetime, and is inversely proportional to the fourth power of the distance from the particle sink. For~this case, the~superfluid velocity increase is inversely proportional to the distance from the particle sink. Due to the pumping effect, the~photon density in~the stationary limit remains roughly constant, implying that the speed of sound for the density fluctuations in the fluid is approximately constant. Hence, in~this case, the~hydrodynamic approximation can be safely employed beyond~the effective acoustic event horizon. Beyond~the acoustic event horizon, the~particle sink can have an important influence on the~dynamics of~phonons within a distance of several times the horizon radius.

\section{Analogue Gravitational~Lensing}
\label{sec_lens}

As discussed in the previous section, the~particle sink creates an effective curved spacetime for the phonons. The~metric in Equation~(\ref{eq_metric}) has no explicit dependence on $t$ and $\varphi$. This means that the associated spacetime is invariant under time translation and rotation along the direction defining the angle $\varphi$, recalling that we only have $2+1$ dimensions. Hence, there are two Killing vectors for this metric, one related to the time-translation invariance $(k_1)^{\mu}=(1,0,0)$ and~the other to the rotational symmetry $(k_2)^{\mu}=(0,0,1)$. Note that the~phonons follow a linear dispersion relation in~the low-energy limit, and as such travel as massless particles on the effective spacetime created by the particle sink, as is made clear by Equation~\eqref{Kleinwave}. Under~the eikonal approximation, if~we assume that the background density and the phase vary slowly in space and time at scales of the wavelength and the period of the perturbation, we expect the phonons to follow null geodesics~\cite{Barcelo2011}. Along these null geodesics, we can define the conserved quantities associated with the Killing vectors $k_1$ and $k_2$ as
\vspace{-6pt}
\begin{eqnarray}
  E &=& -(k_1)^{\mu}g_{\mu\nu}\dv{x^{\nu}}{\lambda}=\left(1-\frac{c_0^2}{r^2}\right)\dv{\tau}{\lambda} , \nonumber\\
  L  &=&(k_2)^{\mu}g_{\mu\nu}\dv{dx^{\nu}}{\lambda}=r^2\dv{\varphi}{\lambda} \, ,
\label{conservation}
\end{eqnarray}
where $E$ is the energy, $L$ is the angular momentum, and $\lambda$ denotes an affine parameter for the null~geodesics.

By rewriting these equations we can obtain the dynamical equations for $\tau$ and $\phi$ with respect to the affine parameter: $\lambda$,\begin{eqnarray}
   \dv{\tau}{\lambda} &=& \frac{Er^2}{r^2-c_0^2} \, , \nonumber\\
    \dv{\varphi}{\lambda} &=& \frac{L}{r^2} \, .
\label{dtl}
\end{eqnarray}

Because the phonons are travelling along null geodesics, the~acoustic metric must fulfill the equation
\begin{equation}
  g_{\mu\nu}\dv{x^{\mu}}{\lambda}\dv{x^{\nu}}{\lambda}=0 \ .
\end{equation}
 Using the expression in Equation~(\ref{eq_metric}), we obtain
\begin{equation}
-\left(1-\frac{c_0^2}{r^2}\right)\left(\dv{\tau}{\lambda}\right)^2+\left(1-\frac{c_0^2}{r^2}\right)^{-1}\left(\dv{r}{\lambda}\right)^2 + r^2\left(\dv{\varphi}{\lambda}\right)^2=0.
\label{nulltraj}
\end{equation}

After substituting $\dd \tau/\dd\lambda$ and $\dd\varphi/\dd\lambda$ into the above equation, we finally obtain the radial equation
\begin{equation}\label{dr2}
\left(\dv{r}{\lambda}\right)^2=E^2+\frac{c_0^2L^2}{r^4}-\frac{L^2}{r^2}.
\end{equation}

Hence, through~Equations~\eqref{dtl} and~\eqref{dr2}, the~path of the phonon is completely specified. In~addition, the~radial motion equation can be recast in energy equation form:
\begin{equation}\label{Vpoten}
  \frac{1}{2}\left(\dv{r}{\lambda}\right)^2+V(r)=\frac{1}{2}E^2,
\end{equation}
with an effective potential
\begin{equation}\label{poten}
  V(r)=\frac{L^2}{2r^2}-\frac{c_0^2L^2}{2r^4}.
\end{equation}

The extreme value of the effective potential is determined by $\dd V(r)/\dd r=0$, leading~to
\begin{equation}\label{rmax}
  r_{\textrm{m}}={\sqrt{2}c_0}.
\end{equation}

As can be seen, at~$r_{\textrm{max}}$ the potential has a maximum value $V(r_{m})=L^2/8c_0^2$, after~which it decreases quadratically as the distance from the particle sink increases. If~the energy of the incident phonon is lower than the potential barrier $E^2 < 2V(r_m)$, then the~phonon can be deflected by the potential. In~this manner, phonons with energy larger than the barrier $E^2>2V(r_m)$ can travel through the barrier and be dragged down to the particle sink. From~the critical value $E^2 = 2V(r_m)$, the corresponding critical impact parameter can be determined as $b_m = L/E = 2c_0$. At~this point, the~phonon rotates around the particle sink in an unstable circular~orbit.

When a phonon is deflected by the particle sink, the~turning point is determined by the condition $\dd r/\dd \varphi=0$, resulting in
\begin{equation}\label{rt}
  r_t = b\sqrt{\frac{1}{2}+\frac{\sqrt{(b^2-4c_0^2)}}{2b}},
\end{equation}
where we have used the impact factor $b=L/E$. If~the phonon is directed towards the particle sink with a large impact factor such that $c_0/b$ is a small parameter, it is possible to expand $r_t$ in series of $c_0/b$ up to second order:
\begin{equation}
    r_t = |b| \left[ 1-\frac{1}{2}\left(\frac{c_0}{b}\right)^2+\mathcal{O}\left\lbrace\left(\frac{c_0}{b}\right)^4\right\rbrace\right].
\end{equation}

The general deflection angle of the phonon passing by the acoustic black hole is obtained by the integral
\begin{equation}\label{dphi}
  \Delta\varphi =2\int_{r_t}^{\infty}\dv{\varphi}{r}=2\int_{r_t}^{\infty}\frac{b}{\sqrt{r^4+b^2(c_0^2-r^2)}}\dd r \ .
\end{equation}

In order to solve this integral, we expand the integrand in powers of the parameter $c_0/b$ up to second order, obtaining the following relation for the deflection angle:
\begin{equation}\label{dangle}
  \Delta\varphi=\pi\text{sign}(b)\left(\frac{3 c_0^2}{4b^2}\right)+\mathcal{O}\left\lbrace\left(\frac{c_0}{b}\right)^3\right\rbrace \, .
\end{equation}

As a consequence, in~the absence of the particle sink the~phonon follows a straight line. The~presence of the particle sink leads to an extra deflection angle. This result is similar for the deflection of the phonon in the acoustic vortex spacetime~\cite{Fischer2002}. However, these have different origins. In~Equation~\eqref{dangle}, the~deflection angle is determined by the dissipation parameter $c_0$, while in the rotational case it is provided by the vortex winding number~\cite{Fischer2002}. In~our case, the~deflections of phonons around the sink resemble the light deflection around a black hole, where the deflection angle is determined by the dissipation parameter $c_0$. In general relativity, the~light deflection angle around the black hole is determined by the black-hole mass; thus, in~our analogue system we can define the effective mass of the acoustic black hole as $c_0^2$, as~already mentioned. In~the case of an irrotational vortex with only an azimuthal flow and no radial flow, no acoustic horizon can be created; however, when the azimuthal flow becomes supersonic, the~phonons can be deflected around the irrotational vortex and the deflection angle is related to its winding~number. 

Moreover, wave scattering around a draining bathtub vortex has been extensively studied~\cite{Dolan2011,Torres2019,Dolan2013,Dempsey2016,Torres2018}. Under~the influence of the effective acoustic metric from the draining bathtub vortex, the~waves can display the analogue Aharonov--Bohm effect, superradiance effect, and~orbiting oscillations in the scattering length~\cite{Dolan2011,Dolan2013,Torres2018}. Furthermore, the~null geodesics incident from spatial infinity can be deflected by the draining bathtub vortex, with~the deflection angle determined by the circulation and draining rate of the vortex~\cite{Dolan2013}. In~contrast to our particle sink model, the~draining bathtub vortex is rotating and resembles a rotating acoustic black hole. Although~the draining bathtub vortex model with zero circulation can reduce to the particle sink model, this last model can nonetheless serve as a different approach for analogue gravitational systems~\cite{Liao2019}. As~an analogue of a Schwarzschild black hole, many properties of its properties can be simulated using the particle sink model~\cite{Barcelo2011,Liao2019}. In~our work, we have proposed the simulation of the gravitational lensing effect around a static acoustic black hole employing the particle sink model, which is different from the draining bathtub vortex results~\cite{Dolan2013}.

The phonon deflection phenomenon induced by the particle sink can lead to an interesting convergence effect. After~travelling along opposite paths with respect to the particle sink center, two initially parallel phonon beams with impact parameter $b$ intersect at a distance $l=2f$ from the sink. From~the deflection angle in Equation~\eqref{dangle}, the~focal length $f$ can be obtained as follows:
\begin{equation}
    f = \frac{2}{3\pi}\frac{b^3}{c_0^2}=\frac{2}{3\pi}\frac{b^3}{r_h^2}.
\end{equation}
Therefore, the~particle sink plays the role of an effective phonon~lens.

\begin{figure}[h]
\vspace{0.4cm}
\begin{tikzpicture}
\coordinate(o) at (0,0); \coordinate(v) at (3,0); \coordinate(s) at (7,0); \coordinate(b) at (3,1.5); \coordinate(d) at (5,2.5);
 \filldraw[thick]
 (0,0) circle (1pt) node[align=left, below] {Source}--
 (3,0) circle (1pt) node[align=left, below] {Sink}--
 (7,0) circle (1pt) node[align=left, below] {Observer};
 \filldraw[thick](0.445696,2.45786) circle (1pt) node[align=left, above] {Image source};
 \draw[dashed,thick] (0.445696,2.45786) -- (3,1.5);
  \draw[-stealth,thick] (0,0)--(1.5,0.75); \draw[thick] (1.5,0.75)--(3,1.5);
  \draw[thick] (3,1.5)--(3,0) node[pos=0.5,right] {$b$}; \draw[dashed,thick] (3,1.5)--(5,2.5);\draw[-stealth,thick] (3,1.5)--(5,0.75); \draw[thick] (5,0.75)--(7,0);
  \draw[<->] (3,-0.5)--(7,-0.5) node[pos=0.5,below] {$d_L$};
  \draw[<->] (0,-1)--(7,-1) node[pos=0.5,below] {$d_s$};
  \pic["$\theta_s$", draw=black, angle eccentricity=1.1, angle radius=1.3cm,right]
    {angle=v--o--b};
    \pic["$\theta_E$", draw=black, angle eccentricity=1.1, angle radius=1.3cm,left]
    {angle=b--s--v};
    \pic["$\Delta\phi$", draw=black,  angle eccentricity=1.1, angle radius=1.0cm,right]
    {angle=s--b--d};
 \end{tikzpicture}
\caption{The source is on the left, the~particle sink in the middle acts as a lens, and the observer on the right receives the deflected phonon~signal. }
\label{Fig::Lens}
\end{figure}
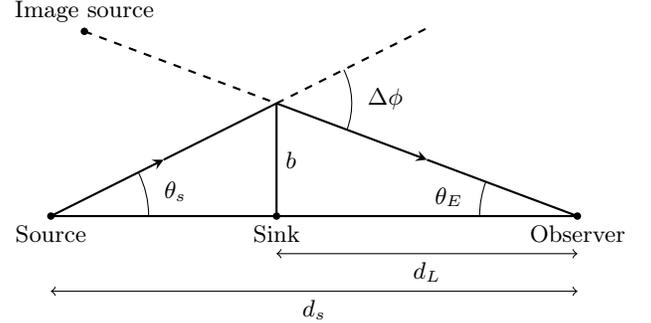

In addition to the phonon convergence effect, similar effects for the phonons can be realized in the case of a particle sink in photonic BECs in~a paradigmatic analogue of gravitational lensing in general relativity. As~depicted in Figure~\ref{Fig::Lens}, if~the source and observer are far away from the particle sink, where $r_h\ll d_L,d_s$, the~phonon beams travel along straight lines and all of the phonon deflections occur around the particle sink. Considering that the angles $\theta_s$ and $\theta_E$ are very small, we obtain
\begin{equation}\label{DL}
  d_L\theta_E=(d_s-d_L)\theta_s=b \, .
\end{equation}

In addition, as can be seen in Figure~\ref{Fig::Lens}, the~deflection angle is equal to
\begin{equation}\label{dphie}
    \Delta \phi = \theta_s+\theta_E,
\end{equation}
where $\Delta\phi= 3 \pi c_0^2/4b^2$ is the deflection~angle.

By combining Equations~\eqref{DL} and~\eqref{dphie}, we obtain the Einstein angle as
\begin{equation}\label{thetaE}
    \theta_E = \left(\frac{3\pi c_0^2(d_s-d_L)}{4d_sd_L^2}\right)^{1/3}.
\end{equation}

In the proposed scenario of photonic BECs, the~sink size is much smaller than the distance between the source, the~particle sink, and~the observer, which leads the particle sink to behave as a thin analogue gravitational lens for the phonons. From~Equation~\eqref{thetaE}, it can be seen that, for a certain $d_L$ satisfying $d_L \ll d_s$, the~Einstein angle $\theta_E$ depends only on the dissipating parameter $c_0$. Comparing this to the case of a Schwarzschild black hole, the parameter $c_0^2$ clearly plays the same role as the black hole mass~\cite{Carroll2019}.
\vspace{-15pt}

In the above discussion, the~analysis of the phonon trajectory deflection is based on the semiclassical approximation; thus, the phonon wavelength should be larger than the healing length. Hence, the~maximum values of the deflection angle (up to second order) induced by the particle sink are determined by
\begin{equation}
	\Delta\varphi = \frac{3\pi}{4}\frac{c_0^2}{\lambda^2} \, ,
\end{equation}
where $\lambda$ is the wave length of the phonons. For~the typical values of the photonic BEC, the~effective mass is $m=6.7\times 10^{-36}$~kg, the~dimensionless interaction strength is of order $\title{g}= (m g)/\hbar^2 \approx 7\times 10^{-4}$, and the particle density is of order $n = 10^{12}$$\sim$$10^{13}{\rm~m}^{-2}$~\cite{Klaers2010b}. Restricting ourselves to the case in which the phonon wave length is in the linear dispersion region with $\lambda=2\xi$, we obtain the deflection angle as follows:
\begin{equation}
	d\phi = \frac{3\pi}{4}(\frac{c_0\hbar}{m c\lambda})^2\approx 0.59^{\circ},
\end{equation}
while the focal length is approximated as
\begin{equation}
	f = \frac{2}{3}\frac{\lambda^3 m^2 c^2}{c_0^2\hbar^2} \approx 20.3~\mu \rm{m} \, .
\end{equation}

In the above estimations, we have restored the physical quantities that we made dimensionless before. Because~the value of the parameter $c_0$ is of order $\mathcal{O}(1)$~\cite{Liao2019,Solnyshkov2019}, we set $c_0=1$. \textcolor{black}{In our calculations, we have used the  quasiclassical phonon scattering theory. Because the transverse size of a phonon beam is limited by its wavelength, in~order to make this approach works the~wavelength has to be less than or equal to the impact factor $b$ of the phonons~\cite{Fischer2002}. Hence, the~minimum focal length is bounded by the smallest available phonon wavelength.} For the dye-based experiments, the~cavity size is about $1.46~~\mu \rm{m}$, which is too small for observing these effects~\cite{Klaers2010b}. However, a~bigger system or~a stronger interaction could be achieved with suitable improvements. According to our calculations, if~the condensed particle density and the inter-particle interaction strength can be increased by one order of magnitude, the~deflection angle can preserve the same value, while the focal length can be reduced to around $2~~\mu \rm{m}$. Therefore, we expect that in the near future our proposal may become feasible for testing the discussed phonon deflection~effects in the lab.

From the experimental perspective, the~analogue gravitational lensing can be implemented in different quantum fluid systems, such as atomic BEC, photonic BEC, and exciton--polariton condensate systems. The~atomic BEC can be relatively easy to manipulate and control experimentally at ultra-cold temperature~\cite{Garay2000,Garay2001}. This makes the atomic BEC very suitable for analogue gravity, while the ultra-cold temperature condition makes it suitable for the study of many quantum effects as well. Until now, analogue phonon Hawking radiation and the related particle entanglement have been realized and observed in atomic BECs~\cite{Steinhauer2014,Steinhauer2016,Kolobov2021,Steinhauer2017}. \textcolor{black}{As discussed before, to~implement the analogue gravitational lensing we need the system to have pumping and dissipation effects in order to compensate each other and create a stable particle sink structure. However, because the atomic BEC does not have intrinsic pumping and dissipation effects, extra experimental techniques are needed to induce pumping and dissipation effects in atomic BEC. In~this case, the~dissipation can be induced by an electron beam shining on the condensate, which can create an effective localized dissipative potential for the condensate~\cite{Barontini2013}. The~pumping effects can be realized by coherent transferring of atoms from a different hyperfine ground state into the condensate~\cite{Doring2009}, although~this would require the atomic BEC to include multiple hyperfine components, which would complicate the problem. Hence, the~atomic BEC is not an ideal platform for analogue gravitational lensing research. Therefore, we concentrate on the quantum fluid system with intrinsic pumping and dissipation effects, such as photonic BECs and exciton--polariton condensates. For~the photonic BEC, its intrinsic pumping and dissipation effects make it feasible to create the stable particle sink structure, and analogue gravitational lensing can be realized in such a system~\cite{Liao2019}.} However, due to the weak inter-particle interaction, as~mentioned before, the~resulting focal length can be very long, which constitute a challenge for current experimental setups. Exciton--polariton condensates can have intrinsic pumping and dissipation effects, and their wave function can be engineered with a versatile all-optical control~\cite{Carusotto2013,Solnyshkov2019}. Because the condensed polaritons have small effective masses and strong inter-particle interactions, the~exciton--polariton condensate can have a large coherence length~\cite{Wertz2010}, and can be controlled with optical potential engineering~\cite{Sanvitto2011}. These characteristics make exciton--polariton condensates very suitable for analogue gravitational lensing~research. 

We next concentrate on how to implement and detect an analogue gravitational lensing experiment in photonic BEC and exciton--polariton condensate. The~implementation of the analogue gravitational lensing experiment first demands the creation of a stable particle sink structure in the condensate. The~particle sink employed here to simulated the black hole can be created in~the photonic BEC and exciton--polariton condensate by etching a hole or placing a scatter in the center of the cavity~\cite{Liao2019,Jacquet2020}. With~such an experimental setup, it would be possible to implement particle pumping with suitable laser beams to create a stationary condensate with a stable particle sink structure. In~order to ensure that the system reaches a stationary state and the particle sink structure becomes stabilized, the~initial pumping should be beyond the stable threshold~\cite{Kneer1998}. When the experiment is initialized, the system can be allowed to evolve to a stationary state and~a stable particle sink structure can be created in the center of the system~\cite{Solnyshkov2019,Jacquet2020}. When the stationary condensate and the stable particle sink are created, the corresponding dissipating parameter $c_0$ can be extracted by~measuring the particle decay rate in the particle sink. With~the stable particle sink structure, it becomes possible to implement and detect the analogue gravitational lensing effect by generating phonons in the condensate and tracking the phonon dynamics around the particle sink. In~order to create and detect phonons in the condensate, we consider Bragg spectroscopy and angle-resolved photoluminescence spectroscopy to be the most suitable methods. For Bragg spectroscopy with~a two-photon Bragg transition, a~well-defined momentum can be imparted to the condensate~\cite{Ozeri2005,Piekarski2021}. For~the photonic BEC and exciton--polariton condensate, they are usually contained in microcavities, and~the microcavities can provide an effective trapping potential for the condensates~\cite{Klaers2010a,Klaers2010b,Carusotto2013}. Under~the influence of the trapping potential, the~imparted momentum cannot be the eigenstate of the condensate, and~will disturb the condensate and create phonon excitations~\cite{Ozeri2005}. Then, by measuring the response of the condensate to Bragg spectroscopy, it is possible to track the phonon dynamics around the particle sink;~the response of the condensate to Bragg transitions of different frequencies provides a spectroscopic measurement of the phonon excitations~\cite{Piekarski2021}. For angle-resolved photoluminescence spectroscopy, phonons can be created in the condensate by applying a weak probe laser beam to the condensate~\cite{Claude2022,Claude2023}. Then, by~measuring the response of the condensate to the probe laser beam, the phonon dynamics around the particle sink can be tracked to determine their trajectories. Moreover, the~phonon spectrum can be extracted from the response of the condensate to the probe laser beam~\cite{Claude2023}. Therefore, by~using Bragg spectroscopy and angle-resolved photoluminescence spectroscopy, the phonon dynamics can be tracked and their trajectories around the particle sink can be determined. By~analyzing the phonon trajectories, we can obtain the corresponding deflection angles and focal lengths of the phonons. Through comparison with the theoretical results, the analogue gravitational lensing effect can then be tested and verified.

\section{Conclusions}
\label{sec_conclusion}

Using a particle sink in photonic BEC as a Schwarzschild black-hole analogue, we have studied the deflection of the phonon trajectories induced by the particle sink at the center of the BEC. This is similar to the light bending effect around a black hole in general relativity. In~the acoustic spacetime approach, the~quasiclassical scattering process of phonons by the particle sink leads to a scattering angle which is quadratic in the dissipating parameter $c_0$. We expect that the initially parallel phonon beams, after~being scattered along opposite side of the particle sink, will~converge at a given distance from the particle sink. This implies that the particle sink can be considered as an effective phonon lens. Analogous to the observation of light bending effects in general relativity, we have discussed the possible thin phonon lens effects induced by the particle sink. Here, the~image photon source location perceived by the observer deviates from the actual source by the Einstein angle $\theta_E$. In~addition, in~the $2+1$ dimensional system there are two image~sources.

We have assumed the hydrodynamic approximation throughout this paper. Consequently, throughout~the process the~system should be kept in a steady state without small-scaled perturbations below the healing length, which is consistent with the requirements of the analogue gravity model~\cite{Barcelo2011}. For~possible experimental implementations, the~photonic BEC in a semiconductor microcavity or dye-based setups might be used, in~particular due to the fact that the semiconductor microcavity setup can have stronger interactions~\cite{Leeuw2016}. 

\section*{Acknowledgements}
This work was supported by the National Institute for the Science and Technology of Quantum Information (INCT-IQ), Grant No.~465469/2014-0, by the National Council for Scientific and Technological Development (CNPq), Grants No~308065/2022-0, and by Coordination of Superior Level Staff Improvement (CAPES). CJ aknowledges 111 Project under Grant No. B2006.


\end{document}